\documentclass[journal]{IEEEtran}
\usepackage{cite}
\usepackage{empheq}
\usepackage{amsthm}
\usepackage{amsmath,amssymb,amsfonts,eqnarray}
\usepackage{graphicx}
\usepackage{textcomp}
\usepackage{xcolor}
\usepackage{acronym}
\usepackage{tabularx}
\usepackage[ruled]{algorithm2e}
\usepackage{algpseudocode}
\usepackage{multirow}
\usepackage{academicons}
\usepackage{scalerel}
\usepackage{tikz}
\usetikzlibrary{svg.path}
\usepackage[table]{xcolor}  
\usepackage{subcaption}
\usepackage{url}
\usepackage{ragged2e}
\usepackage{stfloats}
\usepackage{steinmetz}
\usepackage[symbol]{footmisc}
\usepackage{mathtools}
\usepackage{bm}
\usepackage{makecell}
\usepackage{soul}
\setcellgapes{4pt}
\usepackage{enumitem}
\usepackage{cases}
\usepackage{orcidlink}

\def\BibTeX{{\rm B\kern-.05em{\sc i\kern-.025em b}\kern-.08em
    T\kern-.1667em\lower.7ex\hbox{E}\kern-.125emX}}

\usepackage{hyperref}
 \hypersetup{backref=true,       
    pagebackref=true,               
    hyperindex=true,                
    colorlinks=true,                
    breaklinks=true,                
    urlcolor= black,                
    linkcolor= black,                
    bookmarks=true,                 
    bookmarksopen=false,
    filecolor=black,
    citecolor=black,
    linkbordercolor=black
}
\usepackage{siunitx}

\DeclareSIUnit{\bps}{bps}            
\DeclareSIUnit{\dBm}{dBm}            
\DeclareSIUnit{\dB}{dB}              
\DeclareSIUnit{\THz}{THz}            
\DeclareSIUnit{\GHz}{GHz}            
\DeclareSIUnit{\MHz}{MHz}            
\DeclareSIUnit{\km}{km}              
\definecolor{pastelgray}{RGB}{235,235,235}        
\definecolor{pastelpurple}{RGB}{240,230,250}      
\definecolor{pastelblue}{RGB}{220,240,255}        
\definecolor{pastelnavy}{RGB}{200,220,240}        
\begin{document}
\title{Terahertz Inter-Satellite Links: Motivation, Challenges and Opportunities}
\author{Chandan Kumar Sheemar, \textit{Member, IEEE,} Wali Ullah Khan, \textit{Member, IEEE,} Giovanni Iacovelli, \textit{Member, IEEE,} Sourabh Solanki, \textit{Member, IEEE,}  Mert Bayraktar, \textit{Member, IEEE,} Jorge Querol, \textit{Member, IEEE,}\\ Eva Lagunas, \textit{Member, IEEE} and Symeon Chatzinotas, \textit{Fellow, IEEE}
\thanks{C. K. Sheemar, W. U. Khan, G. Iacovelli, M. Bayraktar, J. Querol, E. Lagunas, and S. Chatzinotas are with the Signal Processing and Communications (SIGCOM) Research Group, Interdisciplinary Centre for Security, Reliability and Trust (SnT), University of Luxembourg, 1855 Luxembourg City, Luxembourg (email: name.surname@uni.lu). S. Solanki is with the ECE Department, National Institute of
Technology Warangal, Telangana, 506004, India (ssolanki@nitw.ac.in). }}

\renewcommand{\qedsymbol}{\scalebox{0.75}{$\blacksquare$}}
\newtheorem{assumption}{Assumption}
\newtheorem{remark}{Remark} 

\newtheorem{theorem}{Theorem}
\newtheorem{corollary}{Corollary}
\newtheorem{lemma}{Lemma}

\maketitle

\begin{abstract}
Inter-satellite links (ISLs) are essential to the evolution of next-generation satellite constellations, providing the foundation for low-latency, resilient, and globally scalable connectivity. While low radio-frequency (RF)-based ISLs offer technological maturity, they are increasingly constrained by spectrum scarcity, congestion, and interference. Optical ISLs, on the other hand, deliver unprecedented capacity but demand ultra-precise pointing, suffer from narrow-beam limitations, and are limited to point-to-point links, all of which hinder large-scale deployment, including point-to-multi-point capability. To overcome these limitations, we propose very-high RF terahertz (THz) inter-satellite links (ISLs) as a promising middle-ground solution, merging the ultra-high data rates of optical links with the adaptability, reliability, and relaxed pointing requirements of lower-frequency RF ISLs. However, despite growing interest, research on THz ISLs remains at an early stage, fragmented across isolated studies, and lacking a clear roadmap for practical realization. This paper aims to address this gap by examining the fundamentals of THz ISLs, assessing their potential advantages and key challenges, and identifying the most promising research directions to transform them into a cornerstone of future interconnected mega constellations.
\end{abstract}


\section{Introduction}
\IEEEPARstart{I}{nter-satellite links }  (ISLs) have become a cornerstone of modern satellite communication networks, enabling direct data exchange between satellites without reliance on ground gateways. By providing autonomous space-based connectivity, ISLs reduce latency, enhance network resilience, and extend global coverage, particularly over regions where ground gateways are impractical, such as the poles, oceans, etc. These capabilities make ISLs indispensable for the emerging non-geostationary orbit (NGSO) constellations \cite{kumar20235g}. While most current deployments rely on intra-orbit links, inter-orbit ISLs are increasingly used to relay large volumes of scientific and Earth observation data from low Earth orbit (LEO) satefllites to higher orbits. As mega-constellations scale to thousands of satellites, ISLs will be pivotal in supporting efficient routing, reducing ground infrastructure dependence, and ensuring harmonious operation of the global satellite systems.

Inter-satellite connectivity is currently enabled by two primary technologies: low\footnote{In this work, microwave and millimeter frequencies are classified as low, compared to the THz frequencies} RF links \cite{ma2011study} and free-space optical ISLs (OISLs) \cite{wang2024free}. Low RF-based ISLs offer technological maturity, interoperability with terrestrial systems, and robust performance. However, limited spectrum availability, growing congestion, and wide beamwidths restrict capacity, reduce spectral efficiency, and risk harmful interference with terrestrial networks. By contrast, OISLs have become the preferred solution \cite{chaudhry2021laser}, providing orders-of-magnitude higher data rates, narrow beams that reduce interference, enhanced physical-layer security, and immunity to spectrum scarcity. However, despite these advantages, OISLs suffer from stringent requirements for sub-microradian pointing, acquisition, and tracking (PAT). Further, OISLs are limited to point-to-point links only and rely on mechanical beam-steering mechanisms, which introduce additional drawbacks, driving the quest for an alternative.


\begin{figure*}[!t]
    \centering
    \includegraphics[width=0.8\linewidth]{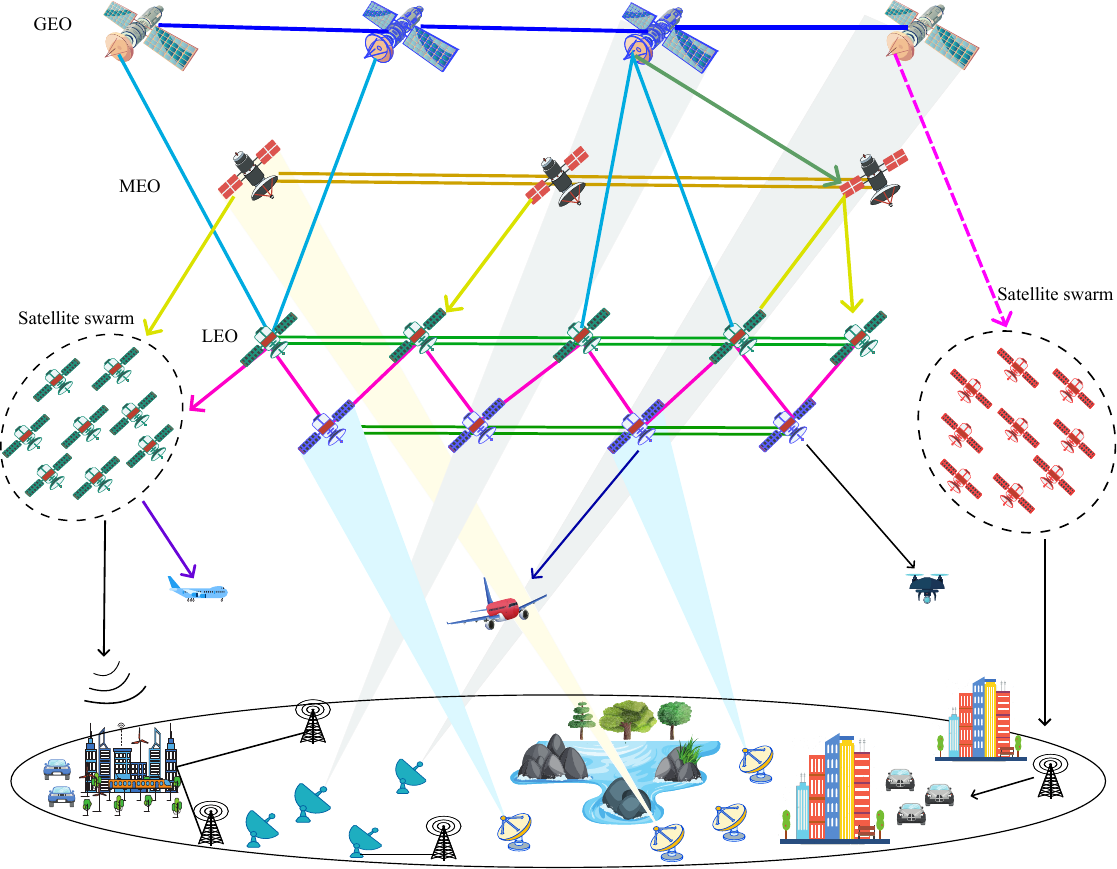}
    \caption{ISLs assisted satellite communication systems.}
    \label{fig1}
\end{figure*}

Recently, terahertz (THz) inter-satellite links have emerged as a compelling alternative, combining the advantages of optical and low-frequency RF systems while addressing their inherent limitations. In \cite{nie2021channel}, the authors developed a channel model for point-to-point THz links
between small satellites, analyzing polarization and frequency diversity along with orbital perturbation effects on beam misalignment. They also concluded that THz ISLs are more robust to environmental factors. In \cite{ding2016analysis}, the authors examined the potential of THz ISLs for ultra-high-speed short- and long-range communication, focusing on antenna design under ideal conditions. In \cite{hu2021deterministic}, a deterministic channel estimation method is
introduced to model scattering from thermal multilayer insulation of THz ISLs. The work in \cite{yang2024channel}
investigated orbital angular momentum-based THz ISLs, modelling inter-channel
interference due to beam deviation. In \cite{li2021propagation}, the authors assessed the performance of THz ISLs for LEO satellites using
finite-difference time-domain methods to account for ionospheric plasma effects and pointing errors. In \cite{tekbiyik2022reconfigurable}, scalability for massive LEO swarms is addressed by proposing multi-antenna strategies and reconfigurable intelligent surfaces to overcome size, weight, and power (SWaP) constraints. The impact of
beam misalignment has been evaluated in \cite{lyu2025inter} through reliability metrics such as bit error rate (BER) and outage probability, considering satellite distances and position perturbations. In \cite{torrens2024modeling}, the authors introduced a
framework for modelling cross-link interference in LEO THz ISLs. In \cite{bakhsh2025direct}, cooperative ISLs for satellite clusters were explored to enable cooperative signal detection. Closed-form ergodic capacity approximations for THz non-ideal conditions were derived.

Although recent studies have begun to explore THz ISLs, the research landscape is still in its infancy and, as such, remains highly fragmented. This work aims to foster a deeper understanding of the fundamental principles of THz inter-satellite communication, emphasizing their significance and unique advantages as a transformative technology that bridges the gap between optical and low RF systems by combining their strengths while mitigating their intrinsic limitations. To realize this vision, however, several challenges must be addressed, which are elaborated in detail. The paper outlines such challenges and pinpoints research directions to make resilient THz ISLs a reality for next-generation mega constellations.

\emph{Paper Organization:} The rest of the paper is organized as follows. Sections \ref{sez_2} and \ref{sez_3} presents the preliminaries and motivation and the fundamentals, distinctive characteristics, and novel use cases, respectively. Section \ref{sez_4} presents a novel analysis relevant to their design. Finally, Sections \ref{sez_5} and \ref{sez_6} discuss challenges and future research directions and conclusions, respectively.

\section{Preliminaries and Motivation} \label{sez_2}

\subsection{The role of ISLs in Satellite Systems}
 ISLs serve as the connective fabric of satellite constellations, enabling direct communication among satellites without reliance on ground infrastructure. At the most fundamental level, ISLs allow satellites to exchange control, synchronization, payload data, and data traffic, which is essential for maintaining network integrity and supporting distributed operations. Two primary forms of ISLs are typically distinguished: intra-orbit links, which connect satellites within the same orbital plane, and inter-orbit links, which connect satellites across different planes or orbital regimes. Together, these links establish a space-based mesh network that enhances coordination and information flow across the constellation. A typical ISLs network containing both the intra-orbit and inter-orbit is shown in Figure \ref{fig1}.
 
Beyond simple data transfer, ISLs play several structural roles in satellite systems. They enable efficient routing, allowing traffic to be forwarded across multiple satellites until an optimal ground station is reached, thereby reducing dependence on any single ground node. ISLs also support load balancing and fault tolerance by distributing traffic dynamically and providing redundant communication paths in case of satellite or gateway failures. 

\subsection{Current ISLs Architectures}

\subsubsection{Low RF ISLs} 
Low RF inter-satellite links are the most established form of connectivity in satellite systems, operating across microwave and/or millimeter-wave frequency bands \cite{chinmayi2016performance}. They rely on well-understood transceiver architectures and standardized spectrum allocations, making them highly interoperable with terrestrial and aerial communication systems. RF ISLs can typically be realized through directional antennas that provide sufficient gain to establish reliable links across the long distances separating satellites. 

Despite their maturity, low RF ISLs face inherent limitations. Spectrum for satellite use is scarce and heavily regulated, leading to growing congestion as the NGSO constellations expand. Their broad beamwidths reduce spectral efficiency, increase susceptibility to interference, and raise the risk of unintended emissions toward terrestrial networks. Moreover, data rates also remain limited, making them challenging to scale for high-throughput applications.

\subsubsection{Optical ISLs}

Free-space OISLs have emerged as the preferred solution for high-capacity satellite communication, which overcomes the challenges of low-RF ISLs. By exploiting the narrow beams of laser transmissions, OISLs achieve data rates orders of magnitude higher than RF systems while offering enhanced physical-layer security and resilience against interference. The narrow beamwidth also reduces the likelihood of unintentional interference with terrestrial networks, alleviating concerns associated with spectrum scarcity and congestion. These features make OISLs highly attractive for NGSO mega-constellations, where efficient spectrum utilization and ultra-high throughput are essential.  

Despite their clear advantages, OISLs face significant limitations that hinder their widespread scalability. A primary challenge is the stringent requirement for sub-microradian PAT \cite{wang2024free}, which is difficult to sustain in the presence of satellite vibrations, thermal fluctuations, and rapid orbital motion. Even slight misalignments can result in severe signal attenuation or complete link failure, thereby reducing link availability. These challenges are exacerbated in NGSO constellations, where frequent handovers and dynamic reconfiguration further strain PAT systems and increase latency. Moreover, the ultra-narrow beams of OISLs preclude point-to-multipoint connectivity, necessitating multiple independent optical terminals per satellite to ensure omnidirectional coverage \cite{wang2024free}. This requirement significantly increases payload mass, power consumption, and integration complexity. Finally, reliance on mechanical beam-steering mechanisms introduces additional drawbacks, including high cost, mechanical wear, limited responsiveness, and vulnerability to environmental disturbances such as thermal cycling and solar radiation. Collectively, these constraints pose serious barriers to the reliability, cost-efficiency, and scalability of OISLs in large satellite networks.

\subsection{Motivation for THz ISLs}
Compared to low RF ISLs, THz inter-satellite links offer vastly greater bandwidth, higher data rates, and narrower beams that enhance spectral efficiency and spatial reuse. Moreover, unlike terrestrial THz propagation, which suffers from severe molecular absorption, the space environment is virtually absorption-free, allowing interference-free operation and naturally safeguarding terrestrial systems from cross-band interference.

Unlike OISLs, THz links can operate with significantly less stringent pointing and alignment requirements, offering enhanced robustness to satellite vibrations, orbital dynamics, and environmental disturbances. This stability translates into longer link durations between satellites in motion, higher availability, and reduced vulnerability to misalignments, which are among the major challenges in OISLs. THz ISLs inherently support scalability through ultra-massive MIMO and holographic beamforming, enabling simultaneous point-to-multi-point connectivity and dynamic routing, infeasible with the current OISLs.

Furthermore, by replacing mechanical PAT mechanisms with electronic beam steering, THz systems simplify payload design, improve reliability, and enhance responsiveness in dynamic orbital environments. Such electronically controlled beamforming architectures allow fast adaptation to changes in satellite mobility, orientation, and constellation topology, supporting communication across dynamic space network conditions. Collectively, these advantages position THz ISLs as a transformative enabler for next-generation satellite networks, offering a practical path toward robust, scalable, and energy-efficient inter-satellite communication.  
\begin{table*}[t]
\centering
\caption{Comparison of low RF, Optical, and THz ISLs}
\label{tab:ISL_comparison}
\renewcommand{\arraystretch}{1.4}
\begin{tabular}{|>{\columncolor{pastelgray}}p{3cm}|
>{\columncolor{pastelpurple}}p{4cm}|
>{\columncolor{pastelblue}}p{4cm}|
>{\columncolor{pastelnavy}}p{4cm}|}
\hline
\textbf{Feature} & \textbf{Low RF ISLs} & \textbf{OISLs} & \textbf{THz ISLs} \\
\hline
\textbf{Operating Spectrum} & Microwave and millimeter-wave bands (Ka, Q/V, E, W, up to $\sim$90 GHz). & Near-infrared (800–1600 nm), typically using laser. & Sub-THz to THz bands (0.1–10 THz, near-term focus on 90–300 GHz and 300–500 GHz). \\
\hline
\textbf{Spectrum Availability \& Regulation} & Mature allocations by ITU, but congested; interference risk with terrestrial systems. & License-free optical bands; no congestion issues but requires careful safety standards. & Regulation emerging (e.g., FCC Spectrum Horizons); largely underutilized and experimental. \\
\hline
\textbf{Bandwidth Potential} & Tens to hundreds of MHz; scarce spectrum limits capacity. & Up to tens of GHz; effectively unlimited spectrum. & Tens of GHz to hundreds of GHz; ultra-wide contiguous bandwidths feasible in space. \\
\hline
\textbf{Typical Data Rates} & 100 Mbps; few Gbps. & 10–100 Gbps demonstrated; Tbps possible. & Tens of Gbps to multi-Tbps (expected with ultra-massive MIMO). \\
\hline
\textbf{Beamwidth \& Pointing Requirements} & Wide beams (degrees); simple pointing, tolerant to misalignment. & Ultra-narrow beams (microradians); stringent PAT with sub-microradian precision. & Narrow beams (milliradians); less stringent than optical, enabling stable links with electronic steering. \\
\hline
\textbf{Hardware Complexity} & Mature RFICs, compact antennas, high TRL. & Requires precision optics, multiple terminals, complex PAT hardware. & Requires THz front-ends, phased arrays/metasurfaces; integration emerging rapidly. \\
\hline
\textbf{Resilience to Misalignment} & High; wide beams tolerate drift and vibrations. & Very low; misalignment can cause direct link failure. & High; beams narrower than low RF but wider than optical; resilient to small errors due to dynamic adaptability. \\
\hline
\textbf{Interference \& Security} & Susceptible to interference and jamming; limited spatial reuse. & Very low interference due to narrow beams; high physical-layer security. & Low interference with electronic narrow multi-beam agility; improved security vs low RF. \\
\hline
\textbf{Environmental Sensitivity} & Minimal impact in space; ionospheric effects at low bands. & Sensitive to vibrations, thermal effects, solar background light. & Largely resilient; weak plasma dispersion/scattering effects. \\
\hline
\textbf{Scalability in NGSO Mega-Constellations} & Limited; bandwidth scarcity and interference constrain scaling. & Challenged; multiple optical terminals, high mass/power overhead per satellite. & Very High; supports multi-link MIMO, beam-hopping, and electronic beam steering. \\
\hline 
\end{tabular}
\end{table*}

\section{Fundamentals, Distinctive Characteristics and Novel Use Cases} \label{sez_3}
This section outlines the fundamentals and the distinctive characteristics of THz ISLs. Subsequently, novel use cases that are beyond the reach of low RF and OISLs are discussed.

\subsection{Spectrum and Regulation}
 The THz spectrum, typically defined within the range of 0.1–10 THz, has recently attracted growing attention for space communications owing to its vast underutilized bandwidth and favorable propagation properties in space. Unlike terrestrial or space-to-ground applications, where atmospheric attenuation is a dominant factor, inter-satellite environments benefit from the absence of atmospheric absorption, rendering large portions of the THz spectrum highly suitable for ultra-high-capacity links. Regulatory bodies have begun to recognize this potential. For instance, the Federal Communications Commission (FCC) introduced the Spectrum Horizons framework, authorizing experimental licenses between 95 GHz and 3 THz for technology validation and prototyping \cite{FCC2019_THz}. At the same time, early demonstration missions, such as the planned THIS-SAT project, are paving the way toward practical evaluations of THz inter-satellite links in orbit. These developments indicate that THz spectrum usage for inter-satellite communications is transitioning from a conceptual phase toward an experimental and regulatory reality, laying the groundwork for standardized deployment in future satellite networks.

\subsection{THz Propagation in Space}
When analyzing THz ISLs, the starting point is the free-space path loss (FSPL). This loss increases with both the distance between satellites the and operating frequency. At first glance, the higher carrier frequencies of THz systems may seem prohibitive. However, the extremely short wavelengths (sub-millimeter scale) allow for very high antenna gains using relatively compact apertures. This means that with properly designed antennas, the additional FSPL at THz can be effectively compensated, enabling reliable communication across distances of several hundred to several thousand kilometers. Moreover, while molecular absorption at THz frequencies is negligible in the vacuum of space, these signals experience strong attenuation upon entering Earth’s atmosphere due to interactions with atmospheric gases. This pronounced absorption provides a natural isolation mechanism, nullifying the interference with terrestrial wireless systems and effectively shielding ground networks from spaceborne THz transmissions.

A key distinction of THz links compared with optical links lies in beamwidth. Optical beams are extremely narrow, demanding sub-microradian pointing accuracy, whereas THz beams are narrower than RF but still wide enough to tolerate small misalignments. This provides a useful trade-off: they maintain high spatial efficiency while relaxing the stringent pointing, acquisition, and tracking requirements that limit optical systems.

Beyond FSPL, other space-specific factors influence propagation at THz frequencies. For example, the relative motion of satellites introduces Doppler shifts, which scale directly with frequency. At THz, the absolute Doppler shift can be significant, but since orbital trajectories are well known, these effects are highly predictable and can be corrected using modern compensation techniques. Another factor is the influence of plasma environments, such as ionospheric irregularities or spacecraft charging. These effects are far less severe at THz compared to low RF bands, though they can still introduce minor dispersion or scintillation in certain conditions.
\begin{figure*}[!t]
    \centering
    \includegraphics[width=1\linewidth]{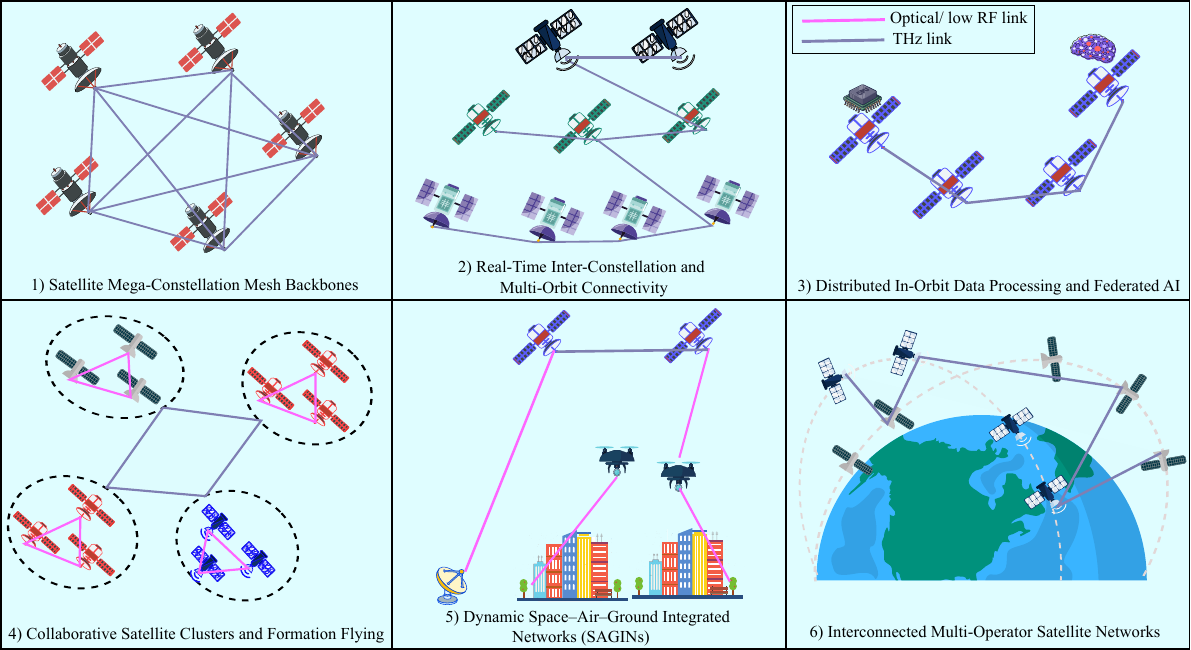}
    \caption{Novel THz-ISLs Enabled Use Cases.}
    \label{figcases}
\end{figure*}

\subsection{System Design Aspects}
 System design for THz ISLs leverages both the high-frequency nature of the spectrum and the unique conditions of space. Antenna systems are central to this design, as electrically large apertures can be realized within relatively small physical footprints. This enables high-gain, narrow-beam transmissions with adjustable beamwidths depending on aperture size. Importantly, while narrower than low RF beams, THz beams remain wider than optical beams, offering an advantageous trade-off between spatial selectivity and tolerance to alignment errors.

Another distinctive feature of THz ISLs lies in their ability to exploit electronic beam steering through phased arrays or reconfigurable metasurfaces. Unlike optical systems, which rely on mechanical steering for pointing, THz antennas can achieve agile, low-latency beam reconfiguration, supporting dynamic connectivity across rapidly moving satellites. This enables multi-link operation, beam hopping, and fine-grained load balancing within large constellations.

From a capacity perspective, the availability of multi-gigahertz to tens-of-gigahertz bandwidths at THz frequencies opens the door to fiber-like throughputs in space. Combined with high antenna gains and spatial multiplexing techniques, THz ISLs can achieve significantly higher aggregate capacity at the network level compared to RF systems, while maintaining more relaxed pointing requirements than optical systems. Thus, THz ISLs occupy a unique middle ground between the robustness of low RF and the capacity of optics, offering a balanced solution for future large-scale satellite networks.

A detailed comparison between the low RF, OISLs and THz ISLs is concisely presented in Table \ref{tab:ISL_comparison}.

\subsection{Novel Use Cases}
In the following subsection, novel use cases that are infeasible with the current low RF (due to limited capacity) and OISLs are discussed, also shown in Fig. \ref{figcases}.

\subsubsection{Satellite Mega-Constellation Mesh Backbones} THz ISLs can enable dense constellations to operate as fully connected mesh networks with simultaneous multi-link connectivity through ultra-massive MIMO and beam-hopping. Unlike OISLs, which are limited to point-to-point links with separate terminals, THz ISLs can dynamically interconnect a massive number of satellites, creating a scalable space-based internet backbone.
\subsubsection{Real-Time Inter-Constellation and Multi-Orbit Connectivity} By supporting multiple concurrent beams with electronic steering, THz ISLs can enable interconnections between LEO, medium Earth orbit (MEO), and geostationary orbit (GEO) systems, as well as between constellations operated by different providers to exchange information, e.g. for interference coordination. Low RF cannot support such high-capacity bridges due to spectral congestion, and OISLs cannot efficiently implement multi-orbit peering with more than four links, which already require four independent transceivers \cite{chaudhry2021laser}.

\subsubsection{Distributed In-Orbit Data Processing and Federated AI} With Tbps-level throughput, THz ISLs make it feasible to interconnect clusters of satellites acting as spaceborne data centers. Satellites can collaboratively process Earth observation data, train AI models in orbit, and distribute results without relying on ground downlinks. Neither low RF (limited capacity) nor OISLs (rigid point-to-point topology) can support the required low-latency, multi-node exchanges.
\subsubsection{Collaborative Satellite Clusters and Formation Flying} Future missions envision distributed telescopes, or synthetic aperture radars (SAR) swarms  or communications satellite swarms operating for a joint purpose. THz ISLs can provide the bandwidth and robustness needed for real-time synchronization, cooperative sensing, and communications across dynamic formations, a capability that the low RF lacks in speed and OISLs cannot sustain due to fragile alignment.
\subsubsection{Dynamic Space–Air–Ground Integrated Networks (SAGINs)} While THz ISLs cannot be directly extended to aerial or terrestrial platforms due to severe molecular absorption in the atmosphere, they can serve as the space backbone of SAGINs. In this role, THz ISLs can provide ultra-high-capacity, low-latency connectivity across dense satellite constellations, which then interface with air and ground segments through low RF or optical gateways. Unlike OISLs, which are constrained to point-to-point links, or low RF, which lacks sufficient bandwidth, THz ISLs can enable the scalable space-layer infrastructure required to support real-time data routing and global backhaul in future SAGIN architectures.
\subsubsection{Interconnected Multi-Operator Satellite Networks} With multiple commercial and governmental constellations emerging, interoperability is a growing requirement. THz ISLs, through multi-beam and beam-hopping capabilities, can allow direct inter-operator peering without duplicating expensive optical terminals or congesting the low RF spectrum. This paves the way for an interconnected \emph{network of networks} in space, something unattainable with current the ISL technologies.

\section{Analysis of THz ISLs} \label{sez_4}
In this section, we provide an analysis of the beam divergence and achievable signal-to-noise (SNR) levels, from which important insights are drawn for system design. These two metrics are particularly critical at THz frequencies, as the highly directional nature of narrow beams directly governs pointing requirements, spatial reuse, and link robustness, while the resulting received SNR determines achievable data rates and communication reliability. Together, beam divergence and SNR capture the fundamental trade-offs between antenna aperture size, alignment accuracy, transmission range, and throughput, making them essential performance indicators for the practical deployment of THz ISLs.

\subsection{Beam Divergence}
Beam divergence in an ISL is primarily governed by the operating wavelength $\lambda$ and the physical aperture diameter $D$. Under diffraction-limited conditions, the full-angle divergence to the first null can be expressed as 
\[
\theta = \frac{2.44 \, \lambda}{D}.
\] 
This relation highlights that, for a fixed aperture, shorter wavelengths result in smaller divergences, while longer wavelengths lead to broader beams. Thus, for the same $D = \SI{10}{\centi\meter}$ aperture, an optical system operating at $\lambda = \SI{1}{\micro\meter}$ yields a divergence on the order of tens of microradians, whereas a THz system at $\lambda = \SI{30}{\micro\meter}$ produces a divergence that is roughly thirty times larger. The parameter dependence is therefore critical in understanding the relative performance of optical and THz ISLs.

\begin{figure}
    \centering
    \includegraphics[width=0.8\linewidth]{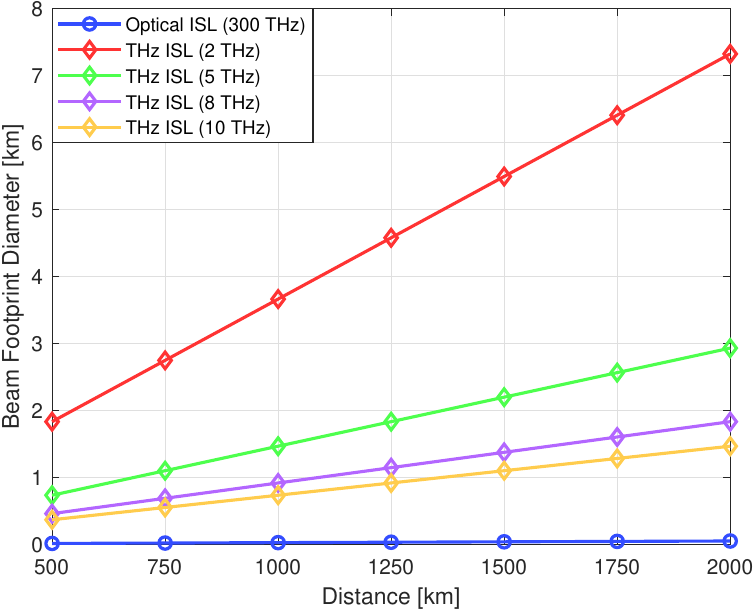}
    \caption{Beam divergence at different THz frequencies.}
    \label{fig:beam}
\end{figure}

As the divergence increases, the beam footprint expands linearly with distance, leading to larger spot sizes at the receiver plane. Compared to optical ISLs, THz beams are significantly wider, although they remain much narrower than conventional low RF beams. This intermediate beamwidth regime has both advantages and drawbacks: while the beams are directional enough to support high-capacity links, they are still wide enough that, at inter-orbital distances, there is a possibility of overlapping footprints and unintended illumination of adjacent satellites. This raises the risk of interference between different orbital shells or constellations, making careful frequency planning, beam pointing, and spatial reuse strategies essential in future THz-based satellite networks. With the aperture of size $D=10~$cm, the divergence of the beam at various THz frequencies is shown in Fig. \ref{fig:beam} as a function of the distance, which shows that at lower THz frequencies the impact of interference towards other ISLs increases. This results drive the quest for proper THz frequency selection, after careful evaluation of the trade-off between interference and pointing requirements relaxation.

\begin{figure}
    \centering
    \includegraphics[width=0.8\linewidth]{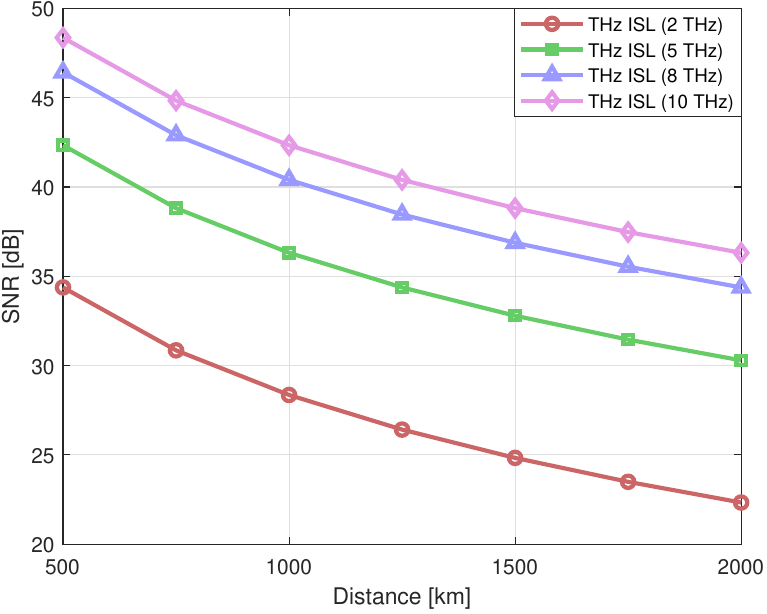}
    \caption{Achievable SNR at different THz frequencies.}
    \label{fig:snr}
\end{figure}
\subsection{Achievable SNR}
For the SNR analysis, we assume a transmit power of $P_t = \SI{1}{\watt}$, a circular aperture antenna of diameter $D = 10~$cm at both the transmitter and receiver, and a receiver noise temperature of $T = \SI{290}{\kelvin}$ with a channel bandwidth of $B = \SI{1}{\giga\hertz}$. The antenna gain is modeled as aperture-limited, given by $
G = \eta \left(\frac{\pi D}{\lambda}\right)^2$,
with $\eta \approx 1$ for the diffraction-limited case, which makes the gain frequency-dependent when the physical aperture is fixed. It is reasonable to assume that antenna gain depends on frequency because the gain of a physical aperture is fundamentally determined by its electrical size relative to the wavelength, meaning that as frequency increases, the same physical antenna becomes electrically larger and focuses energy into a narrower beam. The received power is computed using the Friis transmission equation in free space, neglecting atmospheric attenuation, which is not available in space, thus highlighting the effect of frequency and distance. The resulting SNR is then calculated as 
\[
\text{SNR} = \frac{P_r}{k_B T B},
\] 
where $k_B$ is Boltzmann’s constant. Distances between \SI{500}{\kilo\meter} and \SI{2000}{\kilo\meter} are evaluated, and curves are plotted for various THz carrier frequencies. Fig. \ref{fig:snr} shows that, for a fixed physical aperture, higher THz frequencies yield higher SNR values. This initially appears counterintuitive, since free-space path loss increases with frequency. However, the antenna gain of an aperture is proportional to $ G \propto \left(\frac{D}{\lambda}\right)^{2}$,
so as the wavelength decreases, the antenna becomes electrically larger and more directional, leading to improved energy channelization.

\section{Challenges and Future Research Directions} \label{sez_5}
This section outlines the key challenges and highlights the most promising research directions to enable scalable, high-capacity THz connectivity in the upcoming mega-constellations.

\subsection{Space-Specific Wideband Channel Modelling}
A major barrier to THz ISL deployment is the lack of a comprehensive channel model tailored to the space environment. Existing studies often rely on narrowband LoS assumptions or treat isolated effects such as Doppler or surface scattering, overlooking wideband phenomena intrinsic to THz frequencies. Yet, the vast bandwidth of THz links introduces additional mechanisms, including: frequency-selective fading from structural scattering, polarization mixing, satellite body diffraction, and weak plasma-induced dispersion, that significantly affect system performance. Without accurate models, link budgets, synchronization schemes, and waveform designs remain unreliable, limiting both theoretical and experimental progress.

Developing multi-physics, wideband space-specific channel models is therefore essential. Such models should integrate orbital dynamics, plasma effects, and spacecraft-induced scattering through physics-based ray tracing combined with ephemeris data. Calibration using hardware-in-the-loop emulation and validation via in-orbit experiments (e.g., THIS-SAT) will be key to ensuring realism. To foster community-wide progress, open benchmark datasets and standardized emulators are needed to provide a reproducible foundation for waveform, beam-control, and cross-layer optimization.

\subsection{Beam Management without Mechanical PAT}
Ensuring reliable connectivity in NGSO constellations demands agile and precise beam management under rapid orbital motion, frequent handovers, and structural vibrations. In OISLs, this is achieved through mechanical PAT systems, which, while effective, suffer from latency, wear, and added payload mass. For THz ISLs, the narrow beamwidths render traditional mechanical or codebook-based sweeping approaches inadequate for fast-varying geometries. The challenge intensifies with multi-link operation, where each satellite must simultaneously maintain several concurrent inter-satellite connections, pushing the limits of conventional PAT solutions in terms of responsiveness and scalability.

Future THz systems can rely on predictive, fully electronic beam management to overcome these limitations. Leveraging predictable orbital trajectories, model-based estimators such as Kalman filters or model-predictive controllers can pre-steer beams in anticipation of motion, while inertial measurement unit (IMU) and star-tracker feedback provide real-time refinement. Reconfigurable metasurface and holography based antennas can enable sub-millisecond electronic steering, eliminating the need for mechanical gimbals. Moreover, co-designing beam-hopping, multi-beam scheduling, and routing protocols ensures seamless link continuity during dynamic reconfiguration. The integration of AI-driven control frameworks capable of learning mobility patterns, interference, and traffic dynamics will further enhance adaptability and link reliability across large-scale THz constellations.

\subsection{THz Front-Ends under size weight power (SWaP) and Linearity Constraints}
 THz front-end design for satellites is fundamentally constrained by SWaP limitations. Present THz transceivers suffer from low output power, poor efficiency, high phase noise, and strong nonlinearities in amplifiers and frequency sources, issues further exacerbated by the harsh space environment, where restricted thermal dissipation and radiation exposure degrade performance over time. Conventional digital predistortion is often infeasible at THz due to high complexity and limited effectiveness against memory effects. Moreover, packaging losses at sub-millimeter wavelengths and the scaling of ultra-massive MIMO arrays introduce significant calibration and phase-coherence challenges. Without breakthroughs in front-end efficiency and linearity, the full capacity of THz ISLs cannot be realized.

Advances must center on device, packaging, and architecture co-design to achieve energy-efficient and scalable transceivers. Promising directions include III–V semiconductors (InP, GaN, GaAs) and SiGe BiCMOS/CMOS integration to balance output power and integration density, along with spatial power combining to enhance effective radiated power under thermal limits. On the processing side, hybrid beamforming and low-resolution ADC/DAC architectures can reduce energy demands, while adaptive predistortion and self-calibration can counteract wideband nonlinearities. Finally, thermal-aware circuit design and in-orbit calibration frameworks are essential to maintain long-term stability, enabling practical, high-performance THz payloads for large-scale constellations.

\subsection{Physical Layer Design: Waveforms, Synchronization, and Doppler}
The physical layer at THz faces unprecedented constraints compared to low RF or optical systems. Doppler shifts, which scale linearly with carrier frequency, can reach hundreds of MHz in realistic scenarios, creating severe challenges for synchronization and channel estimation. High phase noise, stemming from oscillators and frequency multipliers at THz, can further degrades conventional multicarrier schemes such as OFDM, leading to inter-carrier interference. At the same time, THz power amplifiers (PA) exhibit strong nonlinearities, making high-peak to power average ratio (PAPR) waveforms inefficient. The enormous bandwidths available also increase pilot overhead and make wideband synchronization computationally intensive.  

Future physical layer designs must be hardware-aware and Doppler-resilient. Promising candidates include filtered-OFDM and tailored low-PAPR multicarrier designs, all of which trade complexity for robustness against phase noise and PA nonlinearities. Synchronization can be improved through ephemeris-assisted Doppler pre-compensation, leveraging orbital predictability to reduce estimation burden. Pilot-efficient channel estimation should be pursued, including compressive sensing methods that exploit sparsity of the THz channel.  
\subsection{Network-Level Scalability: Interference, Scheduling, Routing}
In large-scale NGSO constellations, cross-link interference and uncoordinated resource allocation pose major challenges. Unlike OISLs, whose ultra-narrow beams are naturally interference-free, THz ISLs employ multi-beam operation and aggressive spatial reuse, requiring precise coordination to prevent spectral overlap and degradation. The high density of satellites also demands scalable scheduling and routing mechanisms capable of adapting to rapidly changing topologies, while accounting for limited power, thermal constraints, and the finite number of steerable beams. Without intelligent network-level control, the physical-layer gains of THz links cannot be fully exploited.

Achieving scalability will rely on cross-layer optimization frameworks that integrate interference management, beam scheduling, and routing. At the MAC layer, interference-aware schedulers can exploit predictable orbital geometry using graph-based formulations to allocate beams and bandwidth efficiently. Topology-aware routing and distributed learning approaches can further enhance adaptability and reduce coordination overhead, while beam-hopping enables dynamic capacity allocation in high-demand regions. Incorporating key performance indicators such as energy efficiency, throughput fairness, and link availability into these algorithms will ensure that network-wide scalability is achieved without compromising robustness or efficiency.
\section{Conclusions} \label{sez_6}
This paper presented a comprehensive overview of THz ISLs as a next-generation enabler for scalable, high-capacity satellite networks. It examined the fundamentals and distinctive properties of THz propagation in space, outlined system design aspects, and identified novel use cases beyond the reach of existing RF and optical links. Analytical insights on beam divergence and SNR highlighted the feasibility and design trade-offs of THz operation in orbit. The paper further discussed the key research challenges and proposed corresponding research directions to overcome them. Collectively, these findings establish a roadmap toward realizing practical, robust, and energy-efficient THz ISLs as a cornerstone of future space communication infrastructures.
\bibliographystyle{IEEEtran}
\bibliography{main}

\end{document}